# RFQD - A *DECELERATING* RADIO FREQUENCY QUADRUPOLE FOR THE CERN ANTIPROTON FACILITY

Y. Bylinsky[*], A.M. Lombardi, W. Pirkl, CERN, Geneva, Switzerland

*Abstract*

The RFQD is designed to decelerate antiprotons of momentum 100 MeV/c (kinetic energy 5.33 MeV) down to a kinetic energy variable between ~10 keV and 120 keV. Inside the RFQ body, at ground potential, the rf structure of the four-rod type is mounted on insulating supports. It can be biased between ±60 kV dc to achieve the continuous adjustment of the output energy required by the ASACUSA experiment [1] at the CERN Antiproton Decelerator AD. The different parts of the system are described and the present status reported.

## 1 DESCRIPTION OF THE SYSTEM

The RFQ key parameters are presented in Table 1.

Table 1: RFQ key parameters

| Operating frequency | 202.5 MHz |
|---|---|
| Shunt impedance | 13.8 k$\Omega$ |
| Vane voltage | 167 kV |
| Maximum electric field | 33 MV/m |
| Dissipated power | 1.1 MW |
| Minimum aperture | 0.4 cm |
| Vane modulation factor | 2.9 |
| Input energy | 5.33 MeV |
| Output energy with internal dc post-deceleration | 63 keV (RFQ) ± 60 keV (dc) |
| Transverse acceptance | 15$\pi$ mm mrad |
| Energy acceptance | ± 0.9×10$^{-3}$ |
| Decelerating efficiency | 45% |

The main girder serves as a common support for the medium-energy beam transport (MEBT), the RFQ tank and the low-energy beam transport (LEBT), see Fig.1.

The MEBT starts with two diagnostic boxes joined by a beam transformer. They contain retractable scintillator screens, a silicon strip detector and a Faraday cup. A small rf cavity attached to the RFQ tank allows the input beam energy to be varied. A second cavity of the same type is mounted about 6 m upstream (not shown) to bunch the beam.

The RFQ tank contains the rf structure which consists of a floating "ladder" with 34 rf cells, where the 35 rungs carry the four modulated RFQ electrodes. HV shields are mounted on top and bottom of the ladder which is held in place by 5 ceramic insulators.

The LEBT contains a first solenoid; a first steerer; a diagnostic box with large pumping port and retractable Faraday cup; a second steerer; a SEM wire chamber 95% transparent to the beam that was contributed by ASACUSA; a second solenoid. A measurement line described in paragraph 3 is initially attached instead of the physics experimental apparatus, to assess the RFQ performance.

Three rf amplifier chains power the buncher, energy corrector cavity and RFQ, respectively. Each chain comprises 3 servo loops to stabilise amplitude, phase and resonator tuning. The sheer length of the RFQ (3.55 m) made the usual concept of movable tuners impractical, so the system frequency is determined by the resonance frequency of the RFQ. The error signal generated by the standard tuning criteria is fed to a "motor emulator circuit" which in turn acts on the frequency of the master oscillator. The two low-power resonators use motorized tuners that have to follow the RFQ frequency and absorb also the tuning variations of their respective cavity. The power rating of the RFQ amplifier chain was raised from initially 0.7 MW initially to 2 MW by the addition of a driver stage, to cope with the increased power requirement due to Q-factor lower than expected.

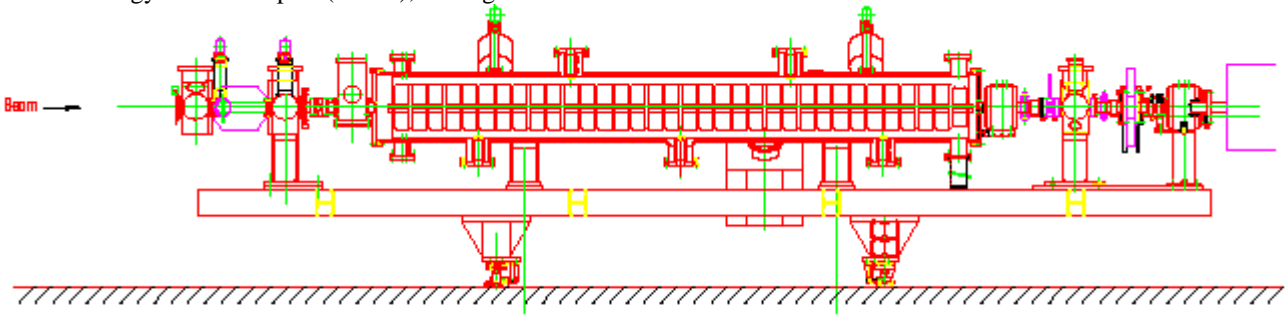

Figure 1: The girder with the MEBT (left), RFQ tank (centre) and LEBT (right)

---

[*] Visitor from INR, Moscow, Russia

## 2 MECHANICS AND VACUUM

**Mechanical layout:** The RFQ tank is a cylinder 3555 mm long, made of 15 mm thick sheet of Cr.Ni.18-10 (Inox 304L). The ladder is milled out of two OFE copper blocks; the electrodes also made of Cu-OFE and are screwed on the ladder together with spiral rf contacts. A dedicated carriage running on demountable rails allows to slide the rf structure into the tank. No special cooling is necessary since the average power dissipation is less than 100 W.

**Vacuum:** Three turbo pumps (250/1400/250 l/s) and three ion pumps (400/800/125 l/s) are mounted in the three girder areas. The LEBT acts as a buffer between the RFQ and the physics apparatus, where there is much higher pressure for the first phase of ASACUSA, but UHV later on. Differential pumping is facilitated by the small output aperture of the RFQ. A pressure of a few times $10^{-8}$ mbar was attained in the RFQ after conditioning with the LEBT end closed. An additional sublimation pump of 2000 l/s is foreseen and bake-out up to 150°C can be applied later if needed.

## 3 BEAM DYNAMICS AND MEASUREMENTS

**Electrode design:** The procedure for the design of the RFQ decelerator has been reported in [2]. The incoming antiproton beam at 5.33 MeV is bunched by a coaxial TEM double-gap resonator, delivering an effective voltage of 47 keV. A corrector cavity, identical to the buncher, is placed at the RFQ entrance to counteract the dc bias of the RFQ ladder, thus preserving the full energy acceptance. The distance between the centre of the buncher and the RFQ determines the longitudinal matching into the decelerating bucket. Magnetic quadrupoles in the upstream MEBT and a conventional radial matching section in the RFQ perform the transverse matching to the time-varying FODO system of the RFQ. The modulated part of the electrodes (3.15 m) decelerates the beam to 63 keV. The modulation factor is maximum at the beginning and decreases as the beam is decelerated in order to keep the maximum field on the vane tip below 1.8 times the Kilpatrick limit, and in order to keep the longitudinal radius of curvature above 5 mm for ease of machining. The beam sizes in the two transverse planes of the decelerated particles are equalised by a carefully designed output matcher. The beam can be further accelerated/decelerated as it exits the RFQ by the dc bias (±60 kV) of the rf structure. In the LEBT a first solenoid focuses the highly divergent beam and makes it almost parallel. A second solenoid at the end focuses the beam into the experimental apparatus.

**Simulation results:** The beam dynamics simulations have been performed with the codes PARMULT, PARMILA and PATH. The expected beam phase space "portraits" at 63 keV are shown in Fig. 2.

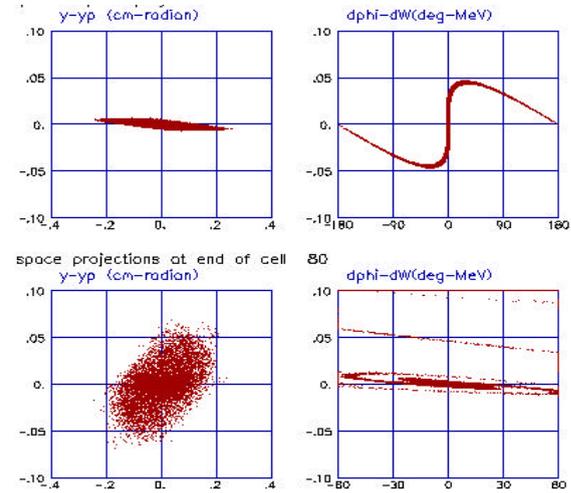

Figure 2: The input (upper) and output (lower) phase space "portraits"

**Measurement line:** In order to assess the RFQ performance, an additional measurement line has been arranged downstream of the LEBT. It comprises a 20° spectrometer magnet, a pair of steerers, a quadrupole triplet, a Faraday cup and a multi-wire profile monitor (see Fig. 3). This set-up allows one to evaluate the beam transmission, output energy, energy dispersion and beam transverse emittance. The whole set-up has been tested with a proton beam of 30 – 80 keV at CERN. Calibration factors for the magnetic field strength were obtained and diagnostic devices were tested at low beam intensity (beam pulse of 50 nA and 5 µs). For transverse emittance evaluation we use a well-known technique based on the profile measurements for different triplet settings. The test results were in good agreement with direct measurements performed by means of a slit – collector emittance measuring device in transport line of the duoplasmatron source.

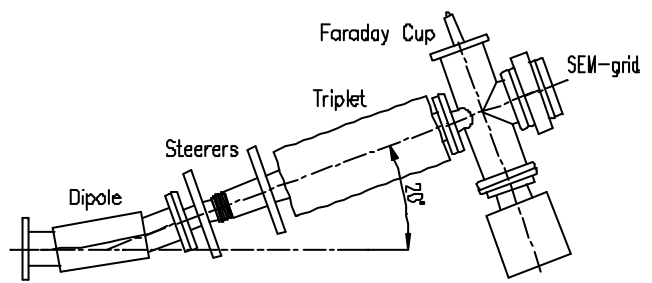

Figure 3: Layout of the measurement line.

## 4 RF ISSUES

**Parasitic modes:** The floating ladder permits the propagation of unwanted parasitic modes that may build up dangerous voltages between the rf structure and the grounded cylinder. These are the TEM(n) and the TE(1,1,n) families that have to be damped or detuned. Low-level tests on a 1/3 scale model started earlier to devise a suitable strategy. Terminating the unwanted modes in capacitively-coupled absorbers led to unreasonably high power losses; therefore (lossless) detuning by 2 pairs of "horns" has been implemented. They consist of open-ended coaxial stubs, close to $\lambda/4$ long, mounted on top and bottom at the two ends of the ladder. A single stub length was found that moves both families sufficiently away from the operating frequency without the need to shift each family separately.

**Power feeding:** Two horizontal coupling loops feed the RFQ at its centre. They are located at a distance of 25 mm from the electrodes to avoid sparking. Their coupling falls short by a factor of about two, hence a transformation circuit is built into the coaxial feeder lines inside the tank. It consists of a low-Z section followed by a folded high-Z section to achieve correct matching. The two power feeder lines are composed of 2 x 4 cables RG218 for high flexibility. A braided shield has been added to the outside of each cable to reduce rf emissions. A 4/1 splitter is provided at each RFQ terminal and an 8/1 splitter at the amplifier.

**Field flattening:** The initially measured resonance frequency, and also the field pattern, differed considerably from the predictions obtained from the usual design programs. This can in part be explained by the substantial length of this RFQ: it becomes highly sensitive to slight local tuning variations due to the large leverage arm of $2.3\lambda$, while the practical limit on the number of mesh points reduces at the same time the precision of a full-length simulation. Movable tuning plates were provisionally mounted in each of the 34 rf cells for the experimental determination of the correct cell heights. The results were then made permanent by slight re-machining of the ladder. Additionally, the mounting of small "fingers" allowed flexible fine-tuning of the assembly. A field flatness of about ±1% was achieved.

**Improvement of the Q-factor:** The initially measured Q-factor was only 3800, about 36% of the theoretical value. Experimental studies on a single full-size cell allowed one to identify some of the loss factors. In particular it was found that the losses in the stainless-steel cylinder were much higher than predicted by simulation. The whole inner cylinder of the RFQ was therefore copper plated, the ladder and the electrodes carefully polished, and all contacts improved. By these measurements a Q-factor of 5100 was finally obtained. In parallel an additional driver stage was included into the rf chain to cover all power requirements with comfortable reserve.

**High power tests:** The RFQ accepted full field after two days of conditioning; the dc bias of 60 kV on the ladder is also reliably sustained. The pulse fault rate is virtually zero; in early tests the equipment ran uninterrupted and unattended over one prolonged weekend.

**Radio Frequency Interference:** The RFQ will have to work in an environment that is very susceptible to electromagnetic perturbations. Emitted rf power from the system has been reduced to about 1 W, thanks to heroic efforts of the RF Group. It was nevertheless decided to enclose the two large amplifiers (and later also the pre-driver) in a Faraday cage to minimise the perturbation to the physics experiments.

## 5 NEXT STEPS

The system has been transported to the "Institut for Fysik" in Aarhus/Denmark where the performance of the RFQ will be assessed using a high-quality proton beam from the Tandem generator. After return to CERN, the RFQ will provide a decelerated antiproton beam for physics from November 2000.

## ACKNOWLEDGEMENTS

The construction of this RFQ would not have been possible without the participation of the PS RF Group and the EST Division. Our thanks go to them and the numerous other persons who actively supported this project.